 \documentclass[necolumn,authoryear]{els-mrw} 

\usepackage{amssymb}
\usepackage{lipsum}
\usepackage{amsmath}

\newcommand{\mpch}{h^{-1}{\rm Mpc}}
\newcommand{\gpch}{h^{-1}{\rm Gpc}}
\newcommand{\msunh}{h^{-1}M_\odot}
\newcommand{\simgt}{\lower.5ex\hbox{$\; \buildrel > \over \sim \;$}}
\newcommand{\simlt}{\lower.5ex\hbox{$\; \buildrel < \over \sim \;$}}

\begin{document}
\chapter{Cosmology with Galaxy Clusters}\label{chap1}

\author[1,2,3]{Hironao Miyatake}
\address[1]{\orgname{Kobayashi-Maskawa Institute for the Origin of Particles and the Universe (KMI), Nagoya University}, \orgaddress{Furo-cho, Chikusa-ward, Nagoya, Aichi, 464-8602, Japan}}
\address[2]{\orgname{Institute for Advanced Research, Nagoya University}, \orgaddress{Furo-cho, Chikusa-ward, Nagoya, Aichi, 464-8601, Japan}}
\address[3]{\orgname{Kavli Institute for the Physics and Mathematics of the Universe (WPI), University of Tokyo}, \orgaddress{5-1-5, Kashiwa-no-ha, Kashiwa, Chiba, 277-8583, Japan}}

\articletag{Chapter Article tagline: update of previous edition,, reprint..}

\maketitle

\begin{glossary}[Glossary]
\term{Cluster abundance} The number density of galaxy clusters as a function of mass at a given redshift, also referred to as the halo mass function. \\
\term{Nuisance parameters} Fitting parameters that are of no interest, but need to be marginalized during the process of model fitting. \\
\term{Stage-III surveys} The third generation surveys, as defined by \cite{Albrecht:2006} which includes DES, KiDS, HSC, ACT, SPT, and SO. \\
\term{Stage-IV surveys} The fourth generation surveys, as defined by
\cite{Albrecht:2006} which includes Euclid, LSST, Roman Space Telescope, and CMB-S4. \\
\term{Thermal Sunyael Zel'dvich effect} The spectral distortion of photons from the CMB due to inverse Compton scattering of CMB photons with hot gas in clusters of galaxies. \citep{Zeldovich:1969,Sunyaev:1972} \\
\term{Weak gravitational lensing} Coherent, subtle shear of distant galaxy images due to the distortion of spacetime caused by foreground structure.\\
\end{glossary}

\begin{glossary}[Nomenclature]
\begin{tabular}{@{}lp{34pc}@{}}
ACT & Atacama Cosmology Telescope \\
CMB & cosmic microwave background \\
CFHT & Canada-France-Hawaii Telescope \\
DES & Dark Energy Survey \\
HSC & Hyper Suprime-Cam \\
HST & Hubble Space Telescope \\
KiDS & Kilo-Degree Survey \\
LSS & large scale structure \\
LSST & Legacy Survey of Space and Time \\
MOR & mass-observable relation \\
O-NIR & optical near-infrared \\
SDSS & Sloan Digital Sky Survey \\
SO & Simons Observatory \\
SPT & South Pole Telescope \\
tSZ & thermal Sunyaev-Zel'dovich \\
WL & weak lensing \\
\end{tabular}
\end{glossary}

\begin{abstract}[Abstract]
We review recent advancements in cosmology with galaxy clusters. Galaxy clusters are the most massive objects in the Universe. Consequently the cluster number density as a function of cluster mass, or cluster abundance, is sensitive to cosmological parameters, particularly the matter density of the Universe $\Omega_{\rm m}$ and the amplitude of matter density fluctuation $\sigma_8$. In this review, we describe the methods used to detect galaxy clusters through optical near-infrared (O-NIR), X-ray, and CMB observations, outlining the advantages and disadvantages of cluster detection through different wavelengths. We describe methods for measuring cluster mass, with a particular focus on calibration by WL measurements. We then discuss how the connection between observables in different wavelengths and cluster abundance can be modeled through a cluster selection function and MOR, and quantify the impact of marginalization of nuisance parameters on cosmological constraints. Finally, we also walk through the recent results of cosmological constraints by cluster abundance with the O-NIR, X-ray, and CMB observations.
\end{abstract}

\begin{BoxTypeA}[chap1:box1]{Key Points}
\begin{itemize}
\item Cluster abundance is sensitive to cosmological parameters ($\Omega_{\rm m}$ and $\sigma_8$). However, there are a number of theoretical and observational challenges.
\item Galaxy clusters can be detected in a variety of wavelengths, including O-NIR, X-ray, and CMB. Each of these detection methods possesses distinct advantages and disadvantages.
\item The estimation of the mass of galaxy clusters is a challenging endeavor. Only WL provides an unbiased mass measurement, as it does not necessitate the use of physical assumptions such as hydrostatic equilibrium or the virial theorem. Recent studies have demonstrated that the application of these physical assumptions can result in a 20-30\% underestimation of the true mass.
\item Linking O-NIR, X-ray, and CMB observations with cluster abundance requires characterizing observational effects, including the cluster selection function and MOR. Marginalizing parameters associated with these effects can degrade the final cosmological constraint.
\item Significant advancements have been made in the field of cluster cosmology over the past two decades. The latest constraints have enabled the measurement of cosmological parameters with a few percent accuracy.
\end{itemize}
\end{BoxTypeA}

\section{Introduction}
\label{sec:introduction}
Galaxy clusters have a long history of use as a cosmological probe. \cite{Zwicky:1933} discovered a `missing mass' around the Coma cluster, which is understood to be dark matter. Measurements of the mass-to-light ratio of galaxy clusters indicated that the matter energy density of the Universe $\Omega_{\rm m}$ is less than 1. The gas-mass fraction $f_{\rm gas}$ at the outskirts of galaxy clusters suggested a low value of $\Omega_{\rm m}$, as reported by \cite{Evrard:1997}.

Recent advances in observational techniques in the O-NIR, X-ray, and CMB domains have elevated cluster abundance to a prominent probe in precision cosmology. The abundance of galaxy clusters can be used to constrain not only $\Omega_{\rm m}$, but also the amplitude of matter density fluctuations $\sigma_8$. Nevertheless, achieving precision measurements of these cosmological parameters presents a number of challenges. One of the most significant challenges is to estimate the mass of galaxy clusters, given that dark matter constitutes a substantial portion of their composition. Additional challenges pertain to the modeling of the relationship between observables derived from O-NIR, X-ray, and CMB data and cluster abundance. Significant advances have been made in the field, enabling partial overcoming of these challenges and resulting in constraints on cosmological parameters with a few percent accuracy.

This review describes the methodology of cosmology with cluster abundance, focusing on the aforementioned challenges. For more detailed discussions on specific points, readers are encouraged to refer to the reviews by \cite{Allen:2011} and \cite{Weinberg:2013}.

\section{Theoretical Background}
\label{sec:theory}
The cluster abundance as a function of halo mass, also known as the halo mass function, is an invaluable tool for advancing our understanding of cosmology. In particular, the exponential tail at the high-mass end is sensitive to cosmological parameters, such as the matter energy density, $\Omega_{\rm m}$, and the standard deviation of the density fluctuations in a sphere with comoving radius $R=8~\mpch$ at present ($z$=0), which is defined as
\begin{equation}
\sigma_8^2 \equiv \left.\int \frac{k^2dk}{2\pi}W^2(kR)P_{\rm lin}(k)\ \right|_{R=8\mpch},
\end{equation}
where $P(k)_{\rm lin}$ is the linear matter power spectrum at present and $W^2(kR)$ is the Fourier transformation of the top-hat window function $W_{R}(x)=3[\sin(x) -x\cos(x)]/x^3$.

The mass function is often assumed to have the following functional form
\begin{equation}
\frac{dn}{d{\rm ln}(M)}=f(\sigma)\frac{\bar{\rho}_{\rm m,0}}{M}\frac{d{\rm ln}(\sigma^{-1})}{d{\rm ln}(M)},
\end{equation}
where $M$ is the cluster mass, $\bar{\rho}_{\rm m,0}$ is the mean mass density of the Universe at present, and $\sigma$ is the rms fluctuations in a sphere with radius $R$ characterized by the cluster mass $M$, defined as
\begin{equation}
\sigma^2 \equiv \left.\int \frac{k^2dk}{2\pi}W^2(kR)P_{\rm lin}(k,z)\ \right|_{R=\left(3M/4\pi\bar{\rho}_{\rm m,0}\right)^{1/3}}.
\end{equation}
\cite{Press:1974} derived the function $f(\sigma)$ through an analytical approach. This was based on the assumption that an object is formed through spherical collapse when the linearly growing density fluctuations exceed a critical density, $\delta_c=1.69$. The Press-Schechter formalism is capable of capturing the qualitative behaviour of the mass function. However, it should be noted that the gravitational collapse is not typically spherical and occurs in the non-linear regime. In a subsequent study, \cite{Sheth:1999} extended the model to encompass ellipsoidal collapse and obtained a better fit to N-body simulations with a box size of $\sim 100\ (\mpch)^3$ and $256^3$ particles. \cite{Tinker:2008} further refined the mass function by calibrating the function $f(\sigma)$ motivated by \cite{Sheth:1999} against N-body simulation suites with a box size spanning from $\sim 100\ (\mpch)^3$ to $\sim 1000\ (\mpch)^3$ populated with $512^3$ to $1024^3$ particles, resulting in a fitting function with an accuracy of $\sim 5 \%$. For a more comprehensive review of the mass function, please refer to \cite{Cooray:2002}. Further refinements and extensions of the mass function beyond \cite{Tinker:2008} are still ongoing. For instance, \cite{Nishimichi:2019} developed an `emulator' of the halo mass function, achieving a few percent accuracy with the exception of the high-mass end ($M\simgt10^{14}\msunh$). They interpolated the parameters in the \cite{Tinker:2008} formalism, which was obtained by fitting measurements of the halo mass function in high-resolution (a box size of $1\gpch$ and $2\gpch$ with $2048^3$ particles) N-body simulation suites with different cosmologies, using a Gaussian Process across cosmological parameters. \cite{Bocquet:2016} considered the impact of the baryonic effects, which pushes matter away from the cluster center through AGN feedback. \cite{Diemer:2020} proposed the mass function as a function of halo mass within a splashback radius, which is defined as the physical boundary of galaxy clusters where incoming particles reach the first apocenter \citep{Diemer:2014,Adhikari:2014,More:2015}. 

Here, we provide a brief overview of the definition of cluster mass. Since galaxy clusters are smoothly connected to other components of the LSS of the Universe, such as filaments and neighboring galaxy clusters or groups, it is not straightforward to define the boundary of galaxy clusters and, hence, cluster masses. One of the physical boundaries of galaxy clusters is the splashback radius, which was proposed about a decade ago. However, measuring the splashback radius for individual galaxy clusters is practically impossible. Instead, the common definition of the cluster mass that has been in use for since several decades ago employs the overdensity of a galaxy cluster. In this instance, the cluster mass is defined such that $M_{\Delta {\rm t}} \equiv 4\pi R_{\Delta {\rm t}}^3 \Delta \rho_{\rm t}/3$, where $R_{\Delta {\rm t}}$ is the cluster-centric radius at which the enclosed mass is equal to $\Delta \rho_{\rm t}$. It is common practice to use $\Delta \rho_{\rm t}=200 \bar{\rho}_{\rm m}(z)$ or $\Delta \rho_{\rm t}=500 \rho_{\rm c}(z)$, where $\bar{\rho}_{\rm m}(z)$ and $\rho_{\rm c}(z)$ is the mean matter density and critical density at redshift $z$, respectively. It should be noted that the radius $R_{\Delta {\rm t}}$ is not comoving, but rather represents physical distance. If $\bar{\rho}_{\rm m,0}$ is used instead of $\bar{\rho}_{\rm m}(z)$, $R_{\Delta {\rm m}}$ represents comoving distance. Another common definition uses the virial radius $R_{\rm vir}$, whereby the cluster mass is defined as $M_{\rm vir} \equiv 4\pi R_{\rm vir}^3 \Delta_{\rm c} \rho_{\rm c}(z)/3$. The value of $\Delta_{\rm c}$ is derived from the collapse of a spherical top-hat perturbation, assuming that the cluster has just virialized. \cite{Bryan:1998} provides a fitting formula for $\Delta_{\rm c}$.

\begin{figure}
	\centering 
	\includegraphics[width=\textwidth]{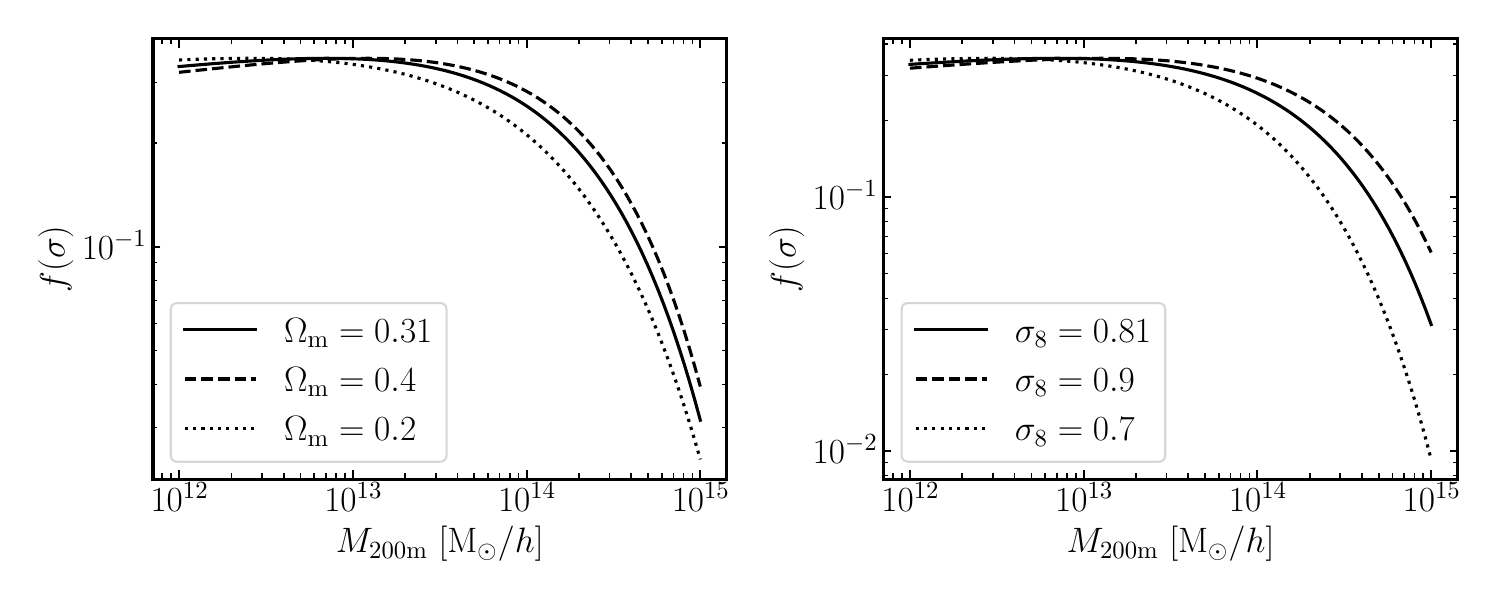}	
	\caption{Behaviour of the function $f(\sigma)$ at $z=0$ with different values of $\Omega_{\rm m}$ and $\sigma_8$, which is computed using the fitting formula in \cite{Tinker:2008}. Cosmological parameters are set to \cite{Planck:2018} except for the large and small $\Omega_{\rm m}$ and $\sigma_8$.} 
	\label{fig:mass_function}%
\end{figure}

Let us see how the mass function behaves against the value of $\Omega_{\rm m}$ and $\sigma_8$. Fig.~\ref{fig:mass_function} shows how the function $f(\sigma)$, which is defined as a function of $M_{\rm 200m}$, depends on these cosmological parameters in the flat-$\Lambda {\rm CDM}$ cosmology. Since $\Omega_{\rm m}$ determines the matter density in the Universe, larger $\Omega_{\rm m}$ results in the larger mass function for a given halo mass at the massive end. Large $\sigma_8$ means the matter distribution in the Universe is more clumpy, and more high-mass clusters are formed.
%


\section{Cluster Samples}
\label{sec:cluster_samples}
The initial step of cluster cosmology is to construct a cluster sample. There are a number of observables for identifying galaxy clusters, each of which has advantages and disadvantages. In this section, we review these observables as well as past and ongoing surveys that are used for constructing cluster samples.

\subsection{Optically-selected Clusters}
\label{sec:optically_selected_clusters}
Optically-selected clusters are identified through the analysis of the concentration of galaxies within a specific region of the sky. This is typically achieved through the utilization of O-NIR imaging data. A mass proxy of an optically-selected cluster catalog is the optical richness, which is defined as the number of member galaxies that satisfy magnitude and color cuts. An optically-selected cluster catalog is often based on survey data covering an area exceeding 100~deg$^2$.

The use of optically-selected clusters offers the advantage of enabling the identification of low-mass halos down to the group scale, thereby providing a larger sample size and more precise statistical constraints on cosmological parameters. Moreover, a reasonable estimate of the redshift can be derived from the data used for cluster identification. In particular, at redshifts $z\simlt1$, early-type galaxies that exhibit a strong Balmer break constitute the majority of cluster members, thereby enabling the acquisition of stable photometric redshifts (photo-$z$s). Although each member galaxy has an estimate of its photometric redshift that is subject to statistical uncertainties in photometry, the statistical uncertainties can be suppressed by averaging over the photo-$z$s of a number of member galaxies. This results in a bias and scatter that are both below 0.005 and 0.01, respectively \citep{Oguri:2018}. 

One disadvantage of an optically-selected cluster sample is that it is susceptible to projection effects, which can introduce bias in cosmological constraints. The projection effects arise from the fact that a cluster finder misidentifies galaxies residing in the LSS other than the cluster (halos and filaments) along the line of sight as cluster members. The projection effects result in an increase in the `observed' cluster richness, leading to the inclusion of low-mass halos in the cluster sample that are not genuinely part of it. Additionally, the projection effects contribute to an enhancement of WL signals at large scales. The impact of these effects on cosmological constraints will be discussed in Section~\ref{sec:current_results}. 

\subsection{X-ray-selected Clusters}
Cluster gas, or the intra-cluster medium (ICM), is quite dilute. Nevertheless, it encompasses the entire massive cluster and possesses a mass that is approximately ten times greater than that of the cluster's member galaxies. Conversely, it is nearly an order of magnitude smaller than the mass of dark matter. Thus, the ICM is the second major component of a galaxy cluster, following dark matter. It constitutes the largest component that can be directly observed through light. The ICM behaves as a fully ionized plasma, which emits a strong X-ray signal through thermal bremsstrahlung. 

The principal advantage of employing X-rays for the purpose of identifying clusters is that they are highly detectable. The emissivity of bremsstrahlung is proportional to the square of the density of the gas, resulting in a significantly stronger X-ray signal from a galaxy cluster than other observables, such as light from member galaxies, which is linearly dependent on the density of components in member galaxies.

One disadvantage is the dimming of X-ray luminosity as a function of redshift. Consequently, the identification of galaxy clusters at high redshifts via X-ray becomes increasingly challenging. Furthermore, previous and current X-ray cluster surveys lack the requisite sensitivity to detect iron lines, thereby preventing the determination of redshift. Consequently, cluster redshifts must be obtained through spectroscopic follow-up or photometric redshifts derived from multi-band images. A further disadvantage is the contamination of active galactic nuclei (AGN), which represents the other significant source of X-ray emission. Typically, X-ray cluster detection necessitates the presence of an extended source. While AGNs are intrinsically point sources, they can be identified as extended sources in dense regions, resulting in the contamination of an X-ray-selected cluster sample.

\subsection{Clusters Selected by the tSZ effect}
The tSZ effect is the spectral distortion of photons from the CMB, which is caused by the inverse Compton scattering of CMB photons with hot gas in clusters of galaxies \citep{Zeldovich:1969,Sunyaev:1972}. Consequently, the intensity at frequencies below (above) 220 GHz is diminished (enhanced).

One advantage of this approach is that a CMB experiment can be designed in such a way that a tSZ cluster sample is not constrained by redshift. This is because tSZ signals are generated with CMB photons as a `backlight.' The limitation of tSZ clusters is not due to redshift; rather, it is determined by the mass of the clusters, which is in turn constrained by the noise level of a CMB experiment. In principle, a mass-threshold sample can be constructed, provided that the beam size of the CMB experiment is sufficiently small to avoid the dilution of tSZ signals. The requisite beam size is contingent upon the size of the cluster. The angular size of a cluster for a given mass decreases as a function of redshift, reaching a minimum at $z\sim1.5$, and then increases. Therefore, if the beam size is smaller than the cluster size at the turnover redshift, any cluster above the specified mass threshold can be detected regardless of redshift. This advantage enables us to investigate the high-redshift Universe in a manner that is not feasible with alternative samples.

One disadvantage of this approach is that it is not possible to determine the redshift of clusters from tSZ signals. Cluster redshifts can be obtained through either spectroscopic follow-up or photometric redshifts. Moreover, the mass limit is generally higher than that of optically-selected clusters or X-ray-selected clusters, particularly at low redshifts. Additionally, at low redshifts, where the typical cluster size is comparable to the scale of CMB fluctuations, clusters may be obscured by CMB fluctuations, resulting in an incomplete cluster sample. Furthermore, CMB fluctuations can be misidentified as clusters, leading to contamination in a cluster sample.


\subsection{Clusters Samples from Past and Ongoing Surveys}
\begin{figure}
	\centering 
	\includegraphics[width=\textwidth]{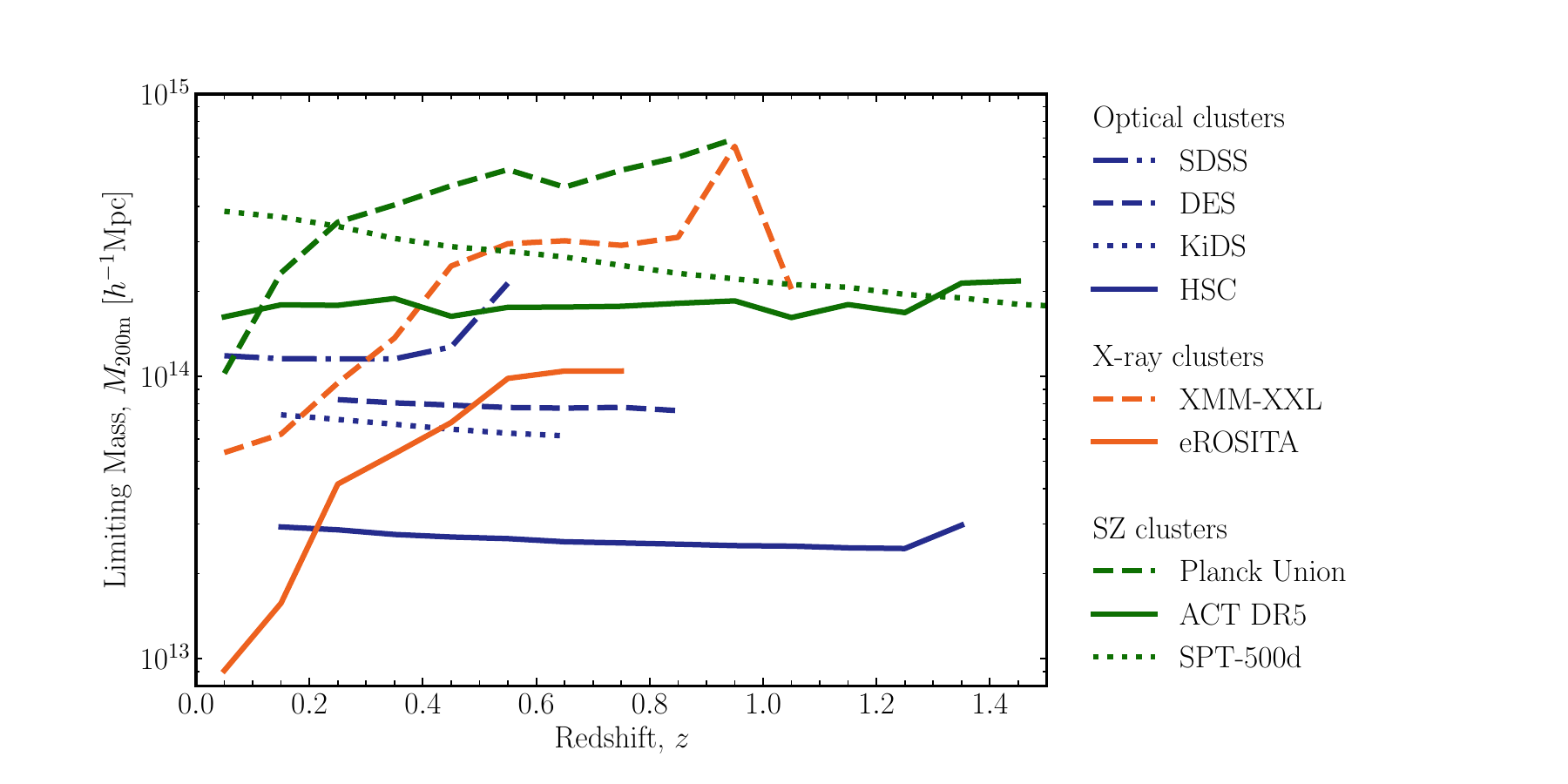}	
	\caption{Limiting mass of optically-selected, X-ray, and SZ clusters detected by past and ongoing surveys as a function of redshift. For detailed explanations of each sample, please see the text.} 
	\label{fig:cluster_samples}%
\end{figure}

Figure~\ref{fig:cluster_samples} illustrates the limiting mass, defined as the lowest cluster mass of a cluster sample, as a function of redshift. This figure presents the results of optical, X-ray, and CMB observations. It should be noted that some of the catalogs lack data regarding cluster mass. In this instance, the mass proxy is converted to cluster mass via a MOR in the literature, which will be described below. In some cases, the catalogs or a MOR does not provide cluster masses in $M_{\rm 200m}$. In these instances, we convert the masses assuming the Navarro-Frenk-White (NFW) profile \citep{Navarro:1996,Navarro:1997} with the concentration-mass relation derived by \cite{Ishiyama:2021}.

\subsubsection{Optically-selected Clusters}
In Fig.~\ref{fig:cluster_samples}, we use the SDSS cluster catalog constructed using the red-sequence Matched-filter Probabilistic Percolation (redMaPPer) algorithm \citep{Rykoff:2014}, which used the spatial concentration of red sequence galaxies. To convert the optical richness to cluster mass, we use the mass-richness scaling relation derived by \cite{Rykoff:2012}. The DES cluster catalog was also constructed by applying the redMaPPer algorithm to the DES-Y1 data \citep{McClintock:2019}. We convert the richness to cluster mass using the scaling relation derived by \cite{McClintock:2019}. The KiDS cluster catalog \citep{Radovich:2017} was constructed using the KiDS ESO-DR2 data \citep{deJong:2015}. They used the Adaptive Matched Identifier of Clustered Objects (AMICO) algorithm, which is based on the optimal filter technique that uses galaxy positions on the sky, magnitudes, and photo-$z$s. The HSC cluster catalog was constructed by applying the cluster finder algorithm called Cluster finding algorithm based on Multi-band Identification of Red-sequence gAlaxies \citep[CAMIRA;][]{Oguri:2014} to the HSC S20A data. CAMIRA is a similar algorithm to redMaPPer in the sense that it utilizes red-sequence galaxies; however, its background subtraction is local, whereas redMaPPer's is global. Note that we apply the richness cut $N>20$ for all the optically-selected cluster catalogs, which normally yields a cluster catalog with high purity (for the KiDS catalog, we applied the cut to $N_{200}$ rather than $N_{500}$).

There are several limiting factors in optical cluster finding. The first limiting factor is the depth of a survey; a shallow survey results in less massive clusters for a given richness cut. This can be seen in Fig.~\ref{fig:cluster_samples}, i.e., the limiting mass decreases as the survey depth increases. Moreover, at the redshift where the magnitudes of member galaxies reach the limiting magnitude of the survey, the cluster sample becomes incomplete, as seen in the SDSS cluster sample. The second limiting factor is the wavelength coverage. As long as we use red sequence galaxies for cluster finding, clusters whose redshift is above the wavelength coverage where the Balmer break falls are undetectable. As seen in Fig.~\ref{fig:cluster_samples}, the redshift range of the HSC cluster catalog is limited to the redshift where the wavelength of the Balmer break becomes longer than the reddest band of HSC ($y$ band). 

Among the Stage-IV surveys, LSST covers 320nm to 1100nm by $ugrzy$ bands, and thus the redshift of clusters is limited to $z\simlt1.3$ as long as red-sequence galaxies are used for cluster finding. However, since LSST will be deeper than HSC, the mass limit will be pushed towards lower mass than that of HSC. In contrast Euclid has a very broad optical band (VIS band), spanning from 550nm to 900nm, and three near-infrared bands ($YJH$ bands), which covers up to 2,020nm. Euclid will use ground-based survey data (LSST, DES, and UNIONS) to cover the optical wavelength with their broad bands. The $YJH$ bands enable the extention of the redshift range of a cluster sample to redshifts $z\simgt1.3$. However, the fraction of red sequence galaxies among cluster member galaxies is expected to become smaller at high redshifts. Consequently, a cluster finder needs to rely on other information such as photo-$z$s. A similar approach should be adopted for Roman, which has seven optical and near-infrared broad bands covering from 480~nm to 2,300~nm.

\subsubsection{X-ray-selected Clusters}
In Fig.~\ref{fig:cluster_samples}, we use the C1 sample from the XMM-XXL cluster catalog \citep{Adami:2018}, which was used for constraining cosmological parameters \citep{Pacaud:2018}. As previously mentioned, X-ray clusters exhibit a decrease in luminosity with increasing redshift, and thus the mass limit increases as a function of redshift, as seen in Fig.~\ref{fig:cluster_samples}. The eROSITA survey observed half of the sky with unprecedented depth, which yielded a cluster sample with much lower limiting mass \citep{Bulbul:2024}. In Fig.~\ref{fig:cluster_samples}, we show clusters used for their cosmology analysis \citep{Ghirardini:2024}. The sample extends down to the group scale at low redshifts.

Unfortunately, the eROSITA survey stopped after half of the survey was completed. Nevertheless, the eROSITA survey is the deepest cluster survey at $z\simlt0.2$ to date. While there are no plans for additional X-ray cluster surveys, the eROSITA X-ray cluster sample will still be valuable at intermediate redshifts ($0.2\simlt z \simlt 0.8$), since it reaches to much lower limiting mass compared to SZ surveys and the theoretical interpretation is generally more straightforward compared to optically-selected clusters since an X-ray cluster sample is less prone to the projection effects.

\subsubsection{tSZ-selected Clusters}
Fig.\ref{fig:cluster_samples} shows the Planck UNION cluster sample \citep{Planck:2016XXVII}. Due to the large beam size ($\sim5$~arcmin), Planck is unable to detect clusters with low mass clusters since these clusters are smeared out. Consequently, the limiting mass of Planck clusters increases with increasing redshift. In contrast, ACT and SPT utilize smaller beams (a few arcmin) compared to Planck, enabling them to detect clusters down to a mass constrained by their noise level ($\sim5-30$~$\mu{\mathrm K}\ {\mathrm arcmin}$), and thus the limiting mass dependence on redshift is moderate. In Fig.~\ref{fig:cluster_samples}, we plot the limiting mass from the ACT DR5 cluster catalog \citep{Hilton:2021} and the SPT-500d cluster catalog \citep{Bleem:2024}, where we applied the signal to noise cut $S/N>5$ for ACT and the cut defined in \cite{Bocquet:2024} for SPT.  

Ongoing and upcoming CMB experiments will push down the limiting mass quite a bit. The ongoing survey by SO will have the capability to identify more than $\sim 16,000$ clusters with a mass greater than $\sim10^{14}\ M_{\odot}$ \citep{Ade:2019}. The forthcoming CMB-S4 survey is expected to achieve a similar or lower mass limit and detect more clusters \citep{Abazajian:2016}.

\section{Cluster Mass Estimation}
The next step in measuring the mass function is estimating the cluster mass. This process is not straightforward because the dominant component of galaxy clusters is dark matter, which does not emit light. Consequently, dark matter cannot be directly observed by a telescope. However, from X-ray or SZ observables, the cluster mass can be estimated assuming hydrostatic equilibrium
\begin{equation}
\label{eq:hse}
\frac{dP_{\rm gas}}{dr} = -\rho_{\rm gas}\frac{GM(<r)}{r^2},
\end{equation}
where $P_{\rm gas}$ and $\rho_{\rm gas}$ is the gas pressure and density, respectively, and $M(<r)$ is the enclosed mass of a cluster within the cluster-centric radius $r$. Nonetheless, in reality, there are non-thermal components, meaning that a system does not reach equilibrium. Given that Eq.~\eqref{eq:hse} does not take into account the non-thermal components, the cluster mass estimated under the assumption of the hydrostatic equilibrium is generally underestimated. Another observable that can be used to estimate cluster mass is the velocity dispersion of member galaxies, which can be measured through spectroscopic observations. Under the assumption of the virial theorem
\begin{equation}
M(<r)=\frac{\kappa r \sigma_{v}^2}{G},
\end{equation}
where $\sigma_{v}$ is the velocity dispersion and $\kappa$ is the coefficient that depends on the density profile which takes $\kappa\sim2$. The dynamical mass again assumes the virial theorem, which is invalid if the system is not completely in equilibrium, and thus cluster mass is generally misestimated. The other observable, WL, offers a unique way to provide an unbiased estimate of galaxy clusters. This method is widely used to estimate a cluster mass or calibrate the X-ray mass and SZ mass. We describe the details of WL and mass calibrations in the following sections. 

\subsection{Weak Lensing}
Gravitational lensing offers a solution to the aforementioned issues by enabling direct measurement of the total mass of a cluster, including dark matter. The bending of light from distant objects, or sources, by foreground massive objects is a consequence of the distortion of space-time around such objects, as predicted by Einstein's general relativity. Consequently, gravitational lensing exhibits sensitivity to both dark matter and baryons.

WL is observed as a subtle distortion, or shear, and magnification or dilation of in the image of source galaxies. Mathematically, weak gravitational lensing is formulated as a coordinate transformation by the Jacobian matrix from the source plane to the image plane
\begin{equation}
{\mathcal A} \equiv
\begin{pmatrix}
1-\kappa+\gamma_1 &\gamma_2\\
\gamma_2 &1-\kappa-\gamma_1\\
\end{pmatrix},
\end{equation}
where $\kappa$ is the convergence and $\gamma$ is the shear, with the limit of $\kappa \ll 1$ \citep[for details, see][]{Bartelmann:2001,Dodelson:2017}. The convergence $\kappa$ is related to the matter distribution of a galaxy cluster as
\begin{equation}
\kappa(\theta)=\frac{\Sigma(D_l\theta)}{\Sigma_{\rm crit}},
\end{equation}
where $\theta$ is the angular distance from the cluster center and $D_l$ is the angular diameter distance from the observer to the cluster. $\Sigma(\theta)$ is the projection of lens matter density profile $\rho(r)$ along the line of sight
\begin{equation}
\Sigma\left(D_l\theta\right) = \int\ dr_{||}\ \rho\left(\sqrt{r_{||}^2+\left(D_l\theta\right)^2}\right),
\end{equation}
and $\Sigma_{\rm crit}$ is the critical surface mass density
\begin{equation}
\Sigma_{\rm cr} = \frac{c^2}{4\pi G}\frac{D_s}{D_l D_{ls}},
\end{equation}
where $D_s$ and $D_{ls}$ are the angular diameter distances from the observer to the source galaxy and from the cluster to the source galaxy. The shear $\gamma$ is written as
\begin{equation}
\gamma(\theta) = \frac{\bar{\Sigma}(<D_l\theta)-\Sigma(D_l\theta)}{\Sigma_{\rm crit}} \equiv \frac{\Delta\Sigma(D_l\theta)}{\Sigma_{\rm crit}},
\end{equation}
where $\bar{\Sigma}(<D_l\theta)$ is the average projected mass density inside $R<D_l\theta$. For a galaxy cluster with a spherical shape, the WL signal exhibits tangential shear with respect to the cluster center\footnote{In practice, shear is frequently used in lieu of magnification. This is due to the fact that shear can be estimated by averaging over a number of galaxies to cancel out their intrinsic shapes. In contrast, to measure magnification, it is necessary to know the properties of a source sample, such as the logarithmic slope of a source sample for a magnitude cut, and hence it is more difficult to control observational systematics.}. In WL measurements, either the shear $\gamma(\theta)$ or the excess surface mass density (ESD) $\Delta\Sigma(R)$, where $R$ is the cluster-centric separation, is used. We use the ESD in the following discussions.

Due to the inherently weak nature of the WL signal, which is typically characterized by an ellipticity of ${\mathcal O(0.01)}$, significantly less than the typical intrinsic galaxy shape of ${\mathcal O(0.3)}$, it becomes necessary to use a substantial number of source galaxies and average out their intrinsic shapes to achieve an accurate measurement. However, it should be noted that the cluster lensing signal remains undetectable for individual less massive clusters, i.e., clusters with mass less than a few $10^{14}$ M$\odot$. Consequently, to enhance the signal-to-noise ratio of cluster lensing measurements, a sample of clusters is `stacked,' i.e., the averaging of a number of cluster lensing signals. Recently, the stacked lensing measurement has been increasingly used in lieu of the lensing measurement of individual clusters. 

Although WL provides unbiased estimates for cluster mass in theory, there are many theoretical and observational systematics to consider. Theoretical systematics include baryonic effects, while observational systematics include imperfect selection of source galaxies. For a thorough review of cluster WL measurements and systematics, readers should refer to \cite{Mandelbaum:2018} and \cite{Umetsu:2020}.

\begin{BoxTypeA}[chap1:box1]{Other methods to estimate cluster mass}
\begin{itemize}
 \item {\bf Strong lensing:} Strong gravitational lensing, which manifests as dramatic phenomena such as giant arcs and multiple images of sources, can also be used to estimate the mass of a cluster. However, it should be noted that strong lensing can only be observed in the inner regions of massive galaxy clusters. Consequently, strong lensing is often used in combination with WL \citep[e.g.,][]{Umetsu:2016}. 
 \item {\bf CMB lensing:} Photons from the CMB are also subject to gravitational lensing. While WL can only measure the foreground mass distribution at redshifts below source galaxies, CMB lensing is sensitive to all cosmic structure at $0<z<1100$. Since CMB lensing is more sensitive to structure at $z\simgt1$, it will be more useful for cosmology with clusters at high-redshift, which can be done in combination with the Stage-VI surveys \citep{Madhavacheril:2017}. 
 \end{itemize}
\end{BoxTypeA}

\subsection{Mass calibration}
\label{sec:mass_calibration}

\begin{figure}
	\centering 
	\includegraphics[width=\textwidth]{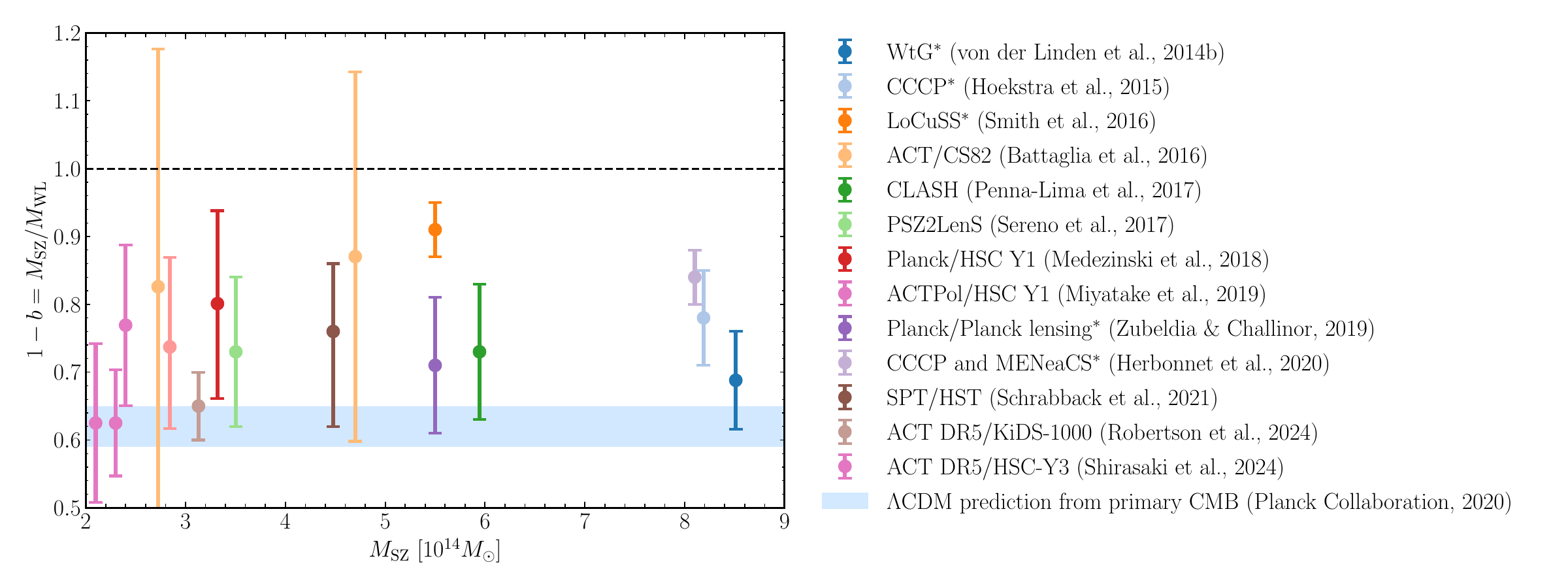}	
	\caption{Hydrostatic mass bias $1-b$ in the literature. The legends with $^*$ denote measurements that need to be corrected for Eddington bias.} 
	\label{fig:mass_bias}%
\end{figure}

Since the cluster mass estimates from X-ray/SZ observables and velocity dispersions are subject to physical assumptions, such as hydrostatic equilibrium and the virial theorem, respectively, the calibration of these observables by mass estimate with WL has been the focus of recent research. In this section, we review mass calibrations performed on X-ray/SZ observables. 

Fig.~\ref{fig:mass_bias} compiles measurements of the so-called hydrostatic mass bias, defined by $1-b\equiv M_{\rm SZ}/M_{\rm WL}$, as a function of $M_{\rm SZ}$, where $M_{\rm SZ}$ is the cluster mass that is derived from the scaling relation between SZ observable and cluster mass derived under the assumption of hydrostatic equilibrium \citep{Arnaud:2010} and $M_{\rm WL}$ is the cluster mass estimated from WL measurements. The Weighing the Giants \citep[WtG;][]{von_der_Linden:2014b} used the 22 clusters in common between the Planck cosmology sample \citep{Planck2014_cluster_sample:2014} and the original WtG sample selected by X-ray \citep{von_der_Linden:2014} and performed WL measurements using Subaru Suprime-Cam and CFHT MegaPrime data, which yielded $1-b=0.688\pm0.072$. The Canadian Cluster Comparison Project \citep[CCCP;][]{Hoekstra:2015} used the 38 clusters in common between the Planck and the original CCCP X-ray cluster sample \citep{Hoekstra:2012} and used the CFHT CFH12k camera and MegaCam, which yielded $1-b=0.76\pm0.05 (stat)\pm0.06 (syst)$. The WtG and CCCP results were used in the Planck cluster cosmology analysis \citep{Planck_2016_cluster_cosmology:2016}, which will be discussed in detail in Section \ref{sec:current_results}. Using Subaru Suprime-Cam, the Local Cluster Substructure Survey (LoCuSS) measured the $1-b$ parameter using 44 clusters in common between the Planck cluster catalog and the original LoCuSS sample \citep[][]{Smith:2016}. \cite{Battaglia:2016} measured the $1-b$ parameter using 9 clusters detected by ACT and CFHT CS82 for WL measurements. They found that the $M_{\rm SZ}$ in the Planck cluster catalog is not corrected for Eddington bias, and thus those $1-b$ measurements and $M_{\rm SZ}$ (indicated with $^*$ in the legend of Fig.~\ref{fig:mass_bias}) should be lowered by 5-10\%. Cluster Lensing And Supernova survey with Hubble \citep[CLASH;][]{Penna-Lima:2017} measured the WL signal around 21 Planck clusters using HST images. Note that they effectively corrected Eddington bias in their analysis. PSZ2LenS \citep{Sereno:2017} measured WL signals around 35 Planck clusters using the CFHTLenS and RCSLenS data, where they already corrected Eddington bias. The Subaru HSC-Y1 data was used for two $1-b$ measurements; one is the measurement with five Planck clusters \citep{Medezinski:2018}, where Eddington bias was corrected, and the other is the measurement with eight ACT clusters \citep{Miyatake:2019} where they found that turning on and off baryonic effects can result in a $1\sigma$ difference in their $1-b$ measurement. Despite the limited number of clusters, the high number density of HSC sources enabled the $1-b$ measurements with the same accuracy as the other measurements. The $1-b$ parameter measured by \cite{Zubeldia:2019} is unique in the sense that they used CMB lensing for estimating cluster mass. \cite{Herbonnet:2020} performed the $1-b$ measurement that used the largest cluster sample at that time. The cluster sample consists of 100 Planck clusters, and they used WL measurements from CCCP and the Multi Epoch Nearby Cluster Survey (MENeaCS). \cite{Schrabback:2021} performed the $1-b$ measurement of clusters selected by SPT-SZ using the HST ACS deep images, which provided the $1-b$ measurement at high redshift ($z_{\rm median}=0.93$). The most recent $1-b$ measurements used $\sim$100 clusters. \cite{Robertson:2024} used 157 ACT clusters and conducted WL measurements using the KiDS-1000 data, while \cite{Shirasaki:2024} used 96 ACT clusters and performed WL measurements with the HSC-Y3 data. \cite{Shirasaki:2024} divided the 96 clusters into three redshift bins, i.e., $0.092<z\leq 0.445$, $0.445<z\leq 0.695$, and $0.695<z\leq 1.180$. However, no significant differences were observed in the $1-b$ measurements. They also find that baryonic effects have the potential to induce a $1\sigma$ shift in the $1-b$ measurement.

Overall, the $1-b$ parameter measurements are around 0.7-0.8 without significant trends as a function of $M_{\rm sz}$ or redshift. The $1-b$ parameter required to reconcile with the $\sigma_8$ constraints by the Planck primary CMB measurements, which is shown in the light-blue shaded region in Fig~\ref{fig:mass_bias}, is smaller than the measured $1-b$ parameter. Thus, from these measurements, the $\sigma_8$ constraint with cluster abundance is expected to be lower than Planck. However, the story would not be that simple as additional factors discussed in the following section can introduce systematic uncertainties.

\section{Cosmological Inference}
\label{sec:cosmological_inference}
In this section, we first introduce an approach commonly used for cosmological inference, which simultaneously models cluster abundance and the stacked lensing signal.

Taking into account the selection function and the MOR, the cluster abundance per steradian is modeled as follows:
\begin{equation}
\label{eq:abundance}
n=\int_{\theta_{\rm obs}\in \theta_{{\rm obs},i}} d\theta_{\rm obs} \int_{z\in z_j} \frac{cD_A^2(z)}{H(z)}dz \int d\ln M \ S(\theta_{\rm obs},z) P(\theta_{\rm obs}|M,z) \frac{dn}{d\ln M},
\end{equation}
where $\theta_{\rm obs}$ is an observable or a derived quantity of a survey used for cluster selection, such as luminosity for an X-ray survey, Compton-$y$ for an SZ survey, and optical richness for an optical survey, $cD_A^2(z)/H(z)$ (consisting of the comoving angular diameter distance $D_A(z)$ and Hubble parameter $H(z)$, e.g., $H(z)=H_0\left[\Omega_{\rm m,0}(1+z)^3+1-\Omega_{\rm m,0}\right]^{1/2}$ for the flat $\Lambda$CDM cosmology) is the cosmological volume, and $dn/d\ln(M)$ is the halo mass function introduced in Section~\ref{sec:theory}. Here, the selection function $S(\theta_{\rm obs},z)$ represents the detection rate of a cluster sample used for cosmological inference, which should be derived for each survey since it depends on the survey depth, detection methods, and observational systematics. The MOR $P(\theta_{\rm obs}|M,z)$ represents the projection from a cluster to mass $M$ at redshift $z$. This projection is inherently probabilistic due to the intrinsic scatter between mass and observables, stemming from the diverse astrophysics that is not fully accounted for solely by cluster mass. Consequently, the MOR is often represented by the mean relation $\ln M=A+B\ln \theta_{\rm obs}$ with scatter between mass and observable $\sigma_{\theta_{\rm obs}|\ln(M)}$, i.e.,
\begin{equation}
\label{eq:scaling_relation}
P(\theta_{\rm obs}|M,z)=\frac{1}{\sqrt{2\pi}\sigma_{\ln\theta_{\rm obs}|\ln M}}\exp{\left[-\frac{\left(\ln M-A-B\ln\theta_{\rm obs}\right)^2}{2\sigma_{\ln\theta_{\rm obs}|\ln M}^2}\right]}.
\end{equation}
A MOR can be extended to use multiple observables. Comprehensive studies of a regression scheme of a multi-variate MOR in the presence of time-evolution, correlated intrinsic scatters, and selection eﬀects (Malmquist / Eddington biases) can be found in a series of CoMaLit studies \cite[Comparing Masses in Literature;][]{Sereno:2015a, Sereno:2015b, Sereno:2015c, Sereno:2015d}, \cite{Sereno:2016}, and \cite{Sereno:2020}. It is important to note that Eq~\eqref{eq:scaling_relation} has two assumptions. First, the mean relation is assumed to have the functional form motivated by a self-similarity of galaxy clusters \citep{Kaiser:1986}, which, in practice, is not the case since clusters are not in equilibrium \citep[see Section 4 in][for details]{Allen:2011}. Second, the log-normal distribution is assumed in Eq.~\eqref{eq:scaling_relation}, but it should be validated by simulations and observations. Furthermore, it is not obvious how the scatter $\sigma_{\ln\theta_{\rm obs}|\ln M}$ depends on cluster properties and the nature of the cluster sample. In the simplest case, a constant value is assumed for the intrinsic scatter, yet it is possible that it varies with cluster mass and redshift. Indeed, \cite{Murata:2018} introduced mass dependence to account for the up-scattering from low-mass clusters due to the projection effects that genuinely exist in an optically-selected cluster sample. In their subsequent analysis, \cite{Murata:2019} further introduced a redshift dependence to the intrinsic scatter. The selection function and scaling relation are integrated over a cluster sample in the $i$-th bin of an observable and the $j$-th redshift bin. This formalism is often called `forward' modeling, since it translates the halo mass function through the MOR for a given halo mass to an observable $P(\theta_{\rm obs}|M,z)$. Conversely, the backward model, i.e., the model that uses the MOR to project an observable to a halo mass $P(M|\theta_{\rm obs},z)$, has been used in the literature \citep[e.g.,][]{Simet:2017,Melchior:2017}.

In the forward modeling method, a stacked WL signal is modeled as
\begin{equation}
\label{eq:stacked_lensing_model}
\left \langle \Delta\Sigma(R)\right \rangle=\frac{1}{n}\int_{\theta_{\rm obs}\in \theta_{{\rm obs},i}} d\theta_{\rm obs} \int_{z\in z_j} \frac{cD_A^2(z)}{H(z)}dz \int d\ln M \ S(\theta_{\rm obs},z) P(\theta_{\rm obs}|M,z) \frac{dn}{d\ln M}\Delta\Sigma(R;M,z),
\end{equation}
where $\Delta\Sigma(R;M,z)$ is the ESD of a density profile for a given halo mass and redshift. As described above, the NFW profile is commonly used for the density profile. In this case, one needs to specify the concentration, which can be computed from a mass-concentration relation. Alternatively, one may opt for alternative profiles, such as
the Einasto profile \citep{Einasto:1965} and a halo profile emulator \citep[e.g.,][]{Nishimichi:2019}. Note that the ESD in Eq.~\eqref{eq:stacked_lensing_model} needs to consider observational systematics such as the off-centering effect \citep[e.g.,][]{Johnston:2007} and photo-$z$ uncertainty \citep[e.g.,][]{Mandelbaum:2008,Nakajima:2012}, and theoretical systematics such as baryonic effects \citep[e.g.,][]{Miyatake:2019,Shirasaki:2024}. Some of these effects are often marginalized over with a reasonable prior.

Next, let us construct a likelihood for cosmological inference. Let us suppose we define $N_{\theta_{\rm obs}}$ bins and $N_z$ bins for the cluster observable and redshift, and $N_R$ bins for the ESD. Then the data vector $\vec{d}$ consists of $n$ for the combinations of $\theta_{\rm obs}$ bins and redshift bins, and $\langle \Delta\Sigma \rangle$ for each of the combinations of $\theta_{\rm obs}$ bins, redshift bins, and $R$ bins. The model vector is defined in a similar manner using Eq.~\eqref{eq:abundance} and Eq.~\eqref{eq:stacked_lensing_model}. Then the likelihood is written as
\begin{equation}
\ln{\mathcal L(\vec{d}\ |\ \vec{{\cal C}},\vec{{\cal N}})}=-\frac{1}{2}\sum_{i,j}\left(\vec{d_i}-\vec{m_i}\right)\left[{\rm Cov}^{-1}\right]_{i,j}\left(\vec{d_j}-\vec{m_j}\right),
\end{equation}
where $\vec{{\cal C}}$ are cosmological parameters, $\vec{{\cal N}}$ are nuisance parameters such as the parameters in the selection function, MOR, and other parameters related to theoretical and observational systematics, and `$\mathrm{Cov}$' is the covariance. Note that the covariance needs to include not only shot noise, such as the Poisson noise in abundance measurements and statistical uncertainties due to intrinsic galaxy shapes in WL measurements, but also sample variance, which arises from observations that are constrained to a finite volume in the Universe \cite[for details, see, e.g.,][]{Oguri:2011}. We then combine priors $P(x)$ to obtain a posterior
\begin{equation}
\ln P(\vec{{\cal C}},\vec{{\cal N}}\ |\ \vec{d})) = \ln{\mathcal L(\vec{d}\ |\ \vec{{\cal C}},\vec{{\cal N}})}+\sum_{x\in(\vec{{\cal C}},\vec{{\cal N})}} \ln P_i(x),
\end{equation}
which needs to be sampled with respect to $(\vec{{\cal C}},\vec{{\cal N}})$ to find the best-fit value, i.e., the maximum a posteriori (MAP) values, and to obtain the posterior distribution.

\begin{figure}
	\centering 
	\includegraphics[width=0.8\textwidth]{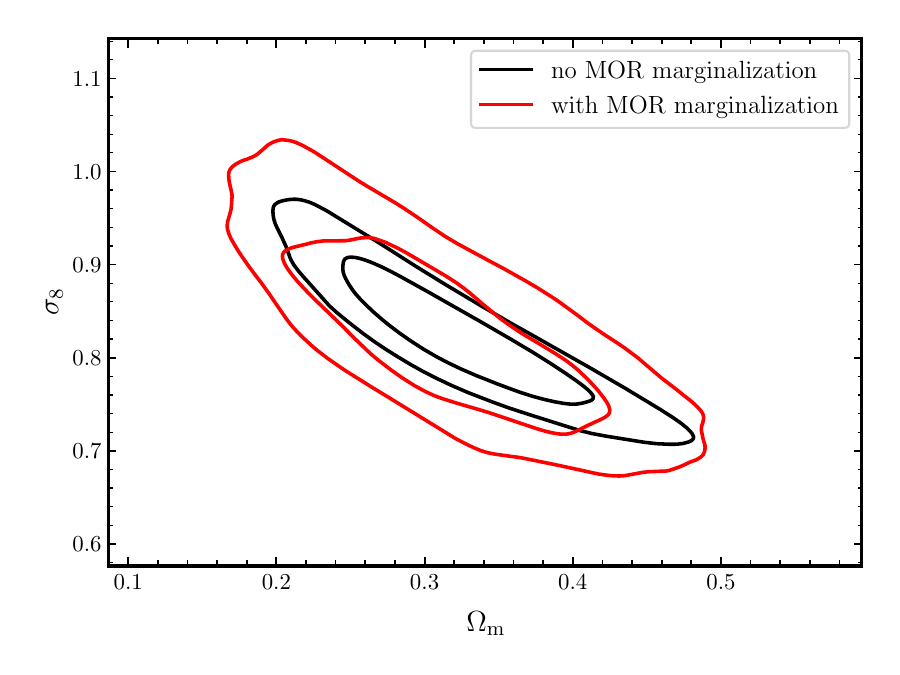}	
	\caption{Forecast of cosmological constraints with a LoVoCCS-like survey. The black (red) line denotes the case without (with) MOR parameter marginalization. The constraining power is significantly degraded when the MOR parameters are marginalized.} 
	\label{fig:forecast}%
\end{figure}

The nuisance parameters need to be marginalized so that systematic uncertainties do not compromise the reliability of a cosmological constraint. For example, since we do not know the exact values in the MOR parameters, they need to be marginalized over the possible uncertainties. Fig.~\ref{fig:forecast} illustrates how a cosmological constraint is degraded when marginalizing the MOR parameters. In Fig.\ref{fig:forecast}, we assume a survey analogous to the ongoing Local Volume Complete Cluster Survey (LoVoCCS) survey \citep{Fu:2022}, which contains 137 galaxy clusters at $0.03<z<0.12$. If we have perfect knowledge of the MOR parameters, i.e., all the MOR parameters are fixed, the statistical uncertainties are reduced by $\sim$40\% in comparison with the scenario in which no knowledge of the MOR parameters is available, that is, marginalizing the MOR parameters with a flat prior. Although it is impossible to obtain perfect knowledge of MOR parameters, a reasonable range of MOR parameters can be obtained, that is to say, informative priors for the MOR parameters can be obtained through simulations or observations. The former can be addressed through hydrodynamical simulations or pasting baryons on N-body simulations, and the latter is actually one of the main purposes of the LoVoCCS survey.

\section{Current Results}
\label{sec:current_results}
\begin{figure}
	\centering 
	\includegraphics[width=0.8\textwidth]{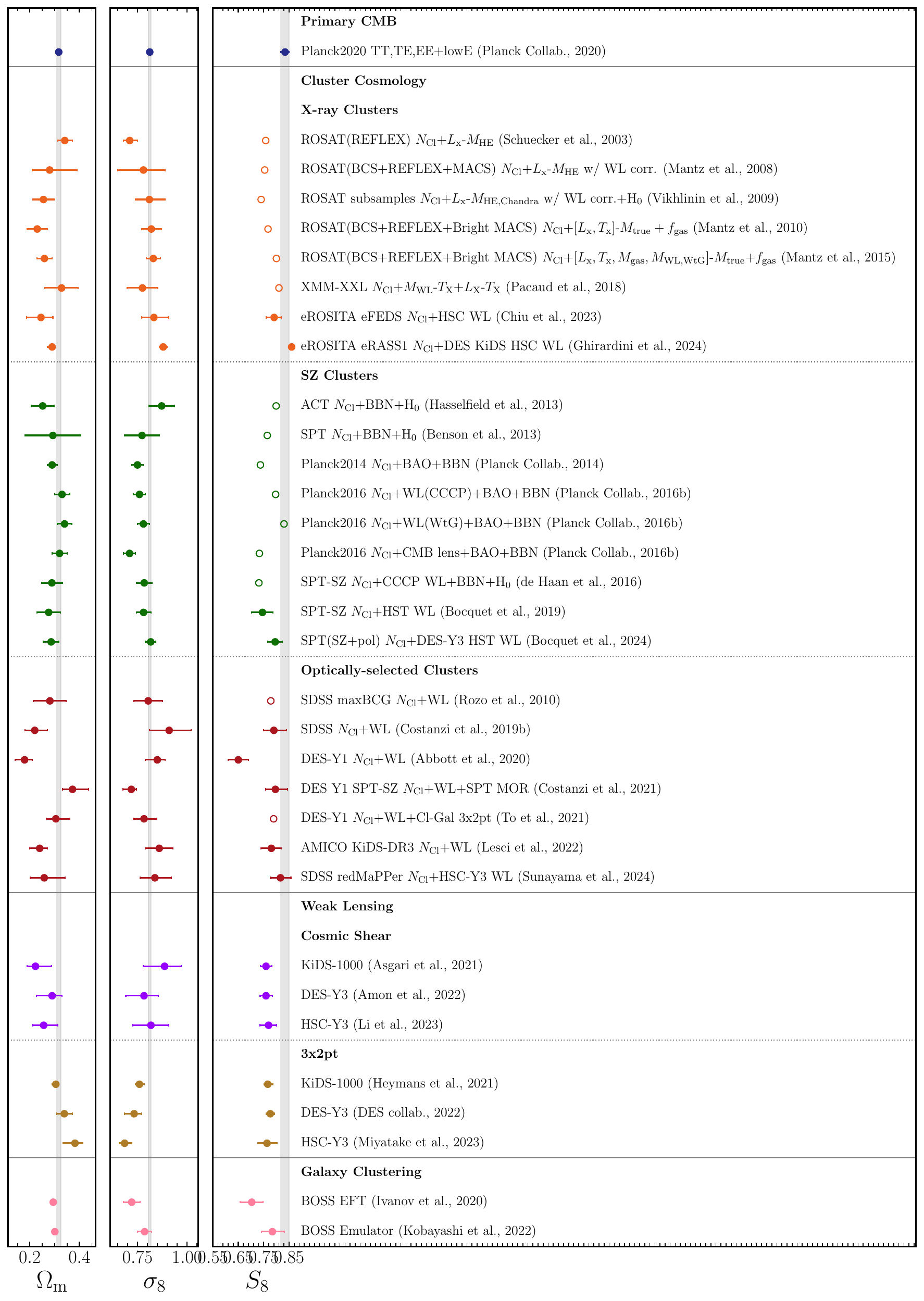}	
	\caption{Constraints on $\Omega_{\rm m}$, $\sigma_8$, $S_8$ with the focus on cluster cosmology. The first block shows the constraint from the Planck primary CMB measurement. The second, third, fourth blocks show cosmological constraints with galaxy clusters selected by X-ray, SZ, and optical imaging data, respectively. The fifth and sixth block show constraints by cosmic shear and the combined probe of cosmic shear, galaxy-galaxy lensing, and galaxy-galaxy clustering, the so-called 3x2pt, respectively. The seventh block show the full-shape analysis of three-dimensional galaxy-galaxy clustering. Computing the error of $S_8$ requires the correlation between $\Omega_{\rm m}$ and $\sigma_8$, which is not available from some of the papers. In this case, we compute the central value of $S_8$ from those of $\Omega_{\rm m}$ and $\sigma_8$, and add it to this figure as open circles.} 
	\label{fig:constraints}%
\end{figure}

Fig.~\ref{fig:constraints} shows the results of cluster cosmology analyses in the last two decades. In this section, we focus on the progress of cosmological constraints on $\Omega_m$, $\sigma_8$, and a derived parameter $S_8\equiv\sigma_8(\Omega_{\rm m}/0.3)^{0.5}$ by cluster abundance selected by different observables. Since the late 2010s, the 2-3$\sigma$ difference in $S_8$ derived from Planck primary CMB measurements and the ones derived from LSS measurements, such as galaxy clusters, WL, and galaxy clustering, has been observed, which is the so-called $S_8$ tension. As a reference the first block of Fig.~\ref{fig:constraints} shows the Planck constraint.

The second block in Figure~\ref{fig:constraints} compiles X-ray cluster cosmology after 2003 \citep[see][for X-ray cluster cosmology before 2003]{Rosati:2002}. The X-ray cluster catalog by ROSAT, the ROSAT All-Sky Survey \citep[RASS;][]{Voges:1999} provided multiple X-ray cluster samples that enabled a number of cosmological analyses with X-ray clusters, such as the ROSAT-ESO Flux-Limited X-ray \citep[REFREX; 447 galaxy clusters above an X-ray flux of $3\times10^{-12}\ {\rm erg} {\rm s}^{-1} {\rm cm}^{-2}$ (0.1 to 2.4 keV) in an area of $\sim$14000~deg$^2$ in the southern sky;][]{Boehringer:2004}, the Massive Cluster Survey \citep[MACS; 124 spectroscopically confirmed clusters at $0.3 < z < 0.7$;][]{Ebeling:2001}, and the Bright Cluster Sample \citep[BCS; 99 clusters at the northern hemisphere, high Galactic latitudes ($|b|\geq20$~deg) and $z\simlt0.3$ with total fluxes between $2.8\times10^{-12}$ and $4.4\times10^{-12}\ {\rm erg} {\rm s}^{-1} {\rm cm}^{-2}$ (0.1-2.4keV);][]{Ebeling:1998,Ebeling:2000}. \cite{Schuecker:2003} constrained $\Omega_m$ and $\sigma_8$ using $\sim$400 galaxy clusters at $0<z<0.3$ in the REFLEX sample and an MOR between X-ray luminosity and cluster mass that is derived assuming hydrostatic equilibrium \citep[hereafter, $L_X$-$M_{\rm HE}$ relation;][]{Reiprich:2002}. \cite{Mantz:2008} added the BCS and MACS samples, the latter of which extended the redshift range up to $z=0.7$, to the REFLEX sample. They calibrated the $L_X$-$M_{\rm HE}$ relation with available WL mass. \cite{Vikhlinin:2009} used 49 low-redshift clusters selected from the ROSAT all-sky survey with the Chandra deep follow-up observations and 37 high-redshift clusters from the ROSAT 400~degree$^2$ serendipitous survey, with a WL-calibrated $L_X$-$M_{\rm HE}$ relation. Although they used only 86 clusters, because of the use of two redshift samples, they obtained stringent constraints. \cite{Mantz:2010} added the gas-mass fraction $f_{\rm gas}$, leveraging the fact that $f_{\rm gas}$ is a constant for the hot, massive clusters \citep{Allen:2008}. \cite{Mantz:2015} combined the WL measurements of 50 clusters to further control systematics in cluster mass. The XMM-XXL survey, which covers $\sim$50~deg$^2$, finally released a new cluster catalog independent from the ROSAT-based catalogs, consisting of 178 galaxy clusters. In the cosmological analysis by \cite{Pacaud:2018}, they used weak lensing mass measured by \cite{Lieu:2016}. Nearly thirty years after the beginning of the ROSAT RASS survey, the eROSITA telescope finally started the eROSITA All-Sky Survey (eRASS). Before starting eRASS, eROSITA conducted the eROSITA Final Equatorial-Depth Survey (eFEDS), which covered $\sim$140~deg$^2$ in the equatorial sky and yielded a catalog that contains 542 galaxy clusters and groups at $0.1<z<1.3$ \citep{Liu:2022}. \cite{Chiu:2023} used 445 clusters to constrain $\Omega_{\rm m}$ and $\sigma_8$ using the WL calibration of 177 clusters with the Subaru HSC data. Then, eROSITA started the all sky survey (eRASS). Using 5,259 clusters detected over an area of 12,791 deg$^2$ \citep{Bulbul:2024} with the cluster masses of $\sim$2,500 clusters calibrated by DES, KiD, and HSC WL measurements, \cite{Ghirardini:2024} yields the tightest constraints on ($\Omega_{\rm m}$,$\sigma_8$) and $S_8$ by X-ray clusters to date, which is consistent with the Planck primary CMB constraint \citep{Planck:2018}.

The third block in Fig.~\ref{fig:constraints} shows the cosmological constraints from SZ-selected clusters. \cite{Hasselfield:2013} used 15 ACT clusters in 270~deg$^2$ and mass calibration with a MOR calibrated against numerical simulations \citep{Bode:2012}. \cite{Benson:2013} used 18 SPT galaxy clusters detected in 178~deg$^2$ and a multi-variable MOR as a function of detection significance and $Y_X$, which is computed from X-ray observables (gas mass $M_{\rm gas}$ and temperature $T_X$) as $Y_X \equiv M_{\rm gas}T_X$. Then the first cluster cosmology results by Planck were released \citep{Planck_2014_cluster_cosmology:2014}, where they used 189 clusters and a MOR between observed SZ flux and hydrostatic mass with a flat prior for the hydrostatic bias parameter $1-b$. The second Planck cluster cosmology results used 439 clusters and reported constraints with different lensing mass calibrations \citep{Planck_2016_cluster_cosmology:2016}, i.e., CCCP \citep{Hoekstra:2015}, WtG \citep{von_der_Linden:2014}, and CMB lensing. Using different mass calibrations resulted in $\simgt 1\sigma$ differences in cosmological constraints, and only WtG yields constraints consistent with Planck primary CMB. After that, SPT expanded their cluster sample. \cite{de_Haan:2016} used 377 clusters selected from the SPT-SZ survey, which covers 2,500~deg$^2$ of the sky, with cluster masses calibrated by CCCP. \cite{Bocquet:2019} used 343 clusters again from the SPT-SZ survey with cluster masses calibrated by WL measurements by the Magellan Telescope and HST, the latter of which enabled calibration of the mass of high-redshift clusters. Their constraint is consistent with \cite{de_Haan:2016}. The latest constraints from SZ-selected clusters is \cite{Bocquet:2024}, where they used 1,005 clusters from the SPT-SZ and SPT-500d survey (5,200 deg$^2$ in total) with the WL mass calibration by DES and HST. They placed the most stringent constraints by SZ-selected clusters to date, which is consistent with the Planck primary CMB measurements. 

The fourth block in Fig~\ref{fig:constraints} shows the cosmological constraints from optically-selected clusters. Since they use optical imaging data, it is natural to use WL for cluster mass calibration. \cite{Rozo:2010} used 9,349 clusters from the SDSS maxBGC cluster catalog \citep{Koester:2007} that covers 7,398~deg$^2$ with cluster mass calibration by SDSS WL measurements. \cite{Costanzi:2019} used the subsample of the SDSS redMaPPer \citep{Rykoff:2014} catalog, which consists of 6,964 clusters over $\sim$10,000~deg$^2$ at $0.1<z<0.3$, with the cluster mass calibration by SDSS WL measurements. This analysis aims to prepare for the DES-Y1 analysis described below by running the DES analysis interference pipeline on the SDSS data. The pipeline includes the model for projection effects which is the contamination to optical richness due to the misidentification of galaxies along the line of-sight towards a galaxy cluster as member galaxies \citep{Costanzi:2019b}. \cite{Abbott:2020} applied the DES analysis pipeline to 6,997 clusters obtained by running the redMaPPer algorithm on the DES-Y1 data that covers $\sim$1,500~deg$^2$ with the redshift range of $0.2<z<0.65$. They used WL measurements run on the DES-Y1 data to estimate cluster masses. They found quite low $\Omega_{\rm m}$ and the constraint on the $\Omega_{\rm m}$-$\sigma_8$ plane has a 2.4$\sigma$ and 5.6$\sigma$ tension with the DES-Y1 3x2pt analysis and the Planck primary CMB measurements. They found that when applying a higher richness cut, the tension is significantly reduced. They concluded that there may be one or more richness-dependent effects not captured by their model. \cite{Costanzi:2021} performed a follow-up analysis using the same DES-Y1 data but adding the data from SPT-SZ. They found their cosmological constraint is consistent with Planck and DES 3x2pt after calibrating their MOR with the SPT-SZ clusters. Since the SPT-SZ clusters correspond to high-richness DES clusters, they concluded that the tension was due to the presence of systematics in the modeling of the stacked weak lensing signal of low-richness clusters. \cite{To:2021} performed another DES-Y1 analysis by adding cluster-cluster clustering, cluster-galaxy lensing, and galaxy-galaxy lensing to cluster abundance and WL. The three clustering and lensing measurements add the ability to constrain their MOR through the mass dependence of halo bias in the cluster WL signal at large scales, which resulted in cosmological constraints consistent with Planck. Cluster cosmology with the other stage-III survey data, i.e., KiDS and HSC, were also reported. \cite{Lesci:2022} used the AMICO KiDS-DR3 catalog that consists of 3,652 galaxy clusters over 377 deg$^2$ at $0.1<z<0.6$ with WL masses, which resulted in a cosmological constraint consistent with Planck. \cite{Sunayama:2024} used 8,379 SDSS redMaPPer clusters over 10,401~deg$^2$ at $0.1<z<0.33$ with HSC-Y3 data to estimate WL masses. They employed a different approach to model projection effects than \cite{Costanzi:2019b}, where they used cluster-clustering signals to constrain the boost in the large-scale WL signals caused by projection effects \citep{Sunayama:2020,Park:2023}. They obtained cosmological constraints as tight as DES and KiDS, and consistent with Planck.

The fifth, sixth, and seventh blocks show cosmological constraints by probes other than galaxy clusters. The fifth block shows constraints by cosmic shear measurements in the Stage-III surveys \citep{Asgari:2021, Amon:2022, Li:2023}. The sixth block shows constraints by 3x2pt which is a combined probe consisting of cosmic shear, galaxy-galaxy lensing, and galaxy-galaxy clustering \citep{Heymans:2021,DES-Y3_3x2pt:2022,Miyatake:2023}. In most cases, these constraints are more stringent compared to cluster cosmology to date and prefer lower $S_8$ compared to Planck. The seventh block shows the so-called full shape analysis of redshift-space three-dimensional galaxy clustering which enables to fully extract cosmological information compared to the traditional methods such as baryon acoustic oscillations (BAO) and redshift-space distortions (RSD). Here, we plot the results from different models, i.e., the effective field theory \citep[EFT;][]{Ivanov:2020} and machine learning-based emulator \citep{Kobayashi:2022}. The former prefers lower $S_8$ compared to Planck, but the latter is consistent with Planck.


\section{Outlook}
In the 2020s, the stage-IV optical and near-infrared surveys, such as Euclid, LSST, Roman, and CMB experiments, such as SO and CMB-S4, will provide data with unprecedented area and depth. These surveys will facilitate the identification of more clusters at $z\simlt1$ and a sizable number of clusters at high redshift. This will enable us to explore beyond $\Lambda$CDM. For instance, the abundance of high-redshift galaxies exhibits enhanced sensitivity to the total neutrino mass  \citep{Ichiki:2012}. The wide coverage of redshift and statistical power will also enable us to constrain other parameters, such as parameters of time-dependent dark energy $(w_0,w_a)$.

The upcoming surveys will significantly reduce the statistical uncertainty, prompting the need to challenge existing modeling of the selection function, MOR, and cluster WL profiles. Since all the cluster finding methods discussed in this review heavily rely on baryons, and cluster astrophysics is not fully understood, it is quite important to understand the behaviour of baryons more deeply through simulations and observations. The limitations of hydrodynamical simulations stem from the assumptions in sub-grid physics. However, by running hydrodynamical simulations with diverse sub-grid physics recipes and varying the parameters of each sub-grid physics model within a reasonable range indicated by observations, we should be able to quantify systematics due to baryons.

In addition, the high source density of the optical and near-infrared stage-IV surveys enables the selection of galaxy clusters from a WL map \citep{Oguri:2024}. In fact, Subaru HSC, which has the highest source density among the stage-III surveys, selected hundreds of clusters from their WL maps \citep[shear-selected clusters;][]{Oguri:2021}, and has successfully placed cosmological constraints \citep{Chen:2024,Chiu:2024}.



\section*{Acknowledgements}
We thank Naomi Robertson and Masato Shirasaki for inputs to make Fig.~\ref{fig:mass_bias}. We use {\tt colossus} \citep{Diemer:2018} to make Fig.~\ref{fig:mass_function}, and {\tt CosmoLike} \citep{Krause:2017} and {\tt GetDist} \citep{Lewis:2019} to make Fig.~\ref{fig:forecast}.

HM is supported by JSPS KAKENHI Grant Numbers JP23H00108, JP23H04005, and JP22K21349 and Tokai Pathways to Global Excellence (T-GEx), part of MEXT Strategic Professional Development Program for Young Researchers.

This work is based on data from eROSITA, the soft X-ray instrument aboard SRG, a joint Russian-German science mission supported by the Russian Space Agency (Roskosmos), in the interests of the Russian Academy of Sciences represented by its Space Research Institute (IKI), and the Deutsches Zentrum für Luft- und Raumfahrt (DLR). The SRG spacecraft was built by Lavochkin Association (NPOL) and its subcontractors and is operated by NPOL with support from the Max Planck Institute for Extraterrestrial Physics (MPE). The development and construction of the eROSITA X-ray instrument was led by MPE, with contributions from the Dr. Karl Remeis Observatory Bamberg \& ECAP (FAU Erlangen-Nuernberg), the University of Hamburg Observatory, the Leibniz Institute for Astrophysics Potsdam (AIP), and the Institute for Astronomy and Astrophysics of the University of Tübingen, with the support of DLR and the Max Planck Society. The Argelander Institute for Astronomy of the University of Bonn and the Ludwig Maximilians Universität Munich also participated in the science preparation for eROSITA. The eROSITA data shown here were processed using the eSASS/NRTA software system developed by the German eROSITA consortium.

Based on data products from observations made with ESO Telescopes at the La Silla Paranal Observatory under programme IDs 177.A-3016, 177.A-3017 and 177.A-3018, and on data products produced by Target/OmegaCEN, INAF-OACN, INAF-OAPD and the KiDS production team, on behalf of the KiDS consortium. OmegaCEN and the KiDS production team acknowledge support by NOVA and NWO-M grants. Members of INAF-OAPD and INAF-OACN also acknowledge the support from the Department of Physics \& Astronomy of the University of Padova, and of the Department of Physics of Univ. Federico II (Naples).

\appendix



\bibliographystyle{elsarticle-harv} 
\bibliography{example}






\end{document}